\documentclass[12pt,preprint]{aastex}

\slugcomment{Accepted to ApJ Letters}
\shortauthors{Howell et al.}
\shorttitle{Polars with Spitzer}

\begin{document}

\title{
First Spitzer Space Telescope Observations of Magnetic 
Cataclysmic Variables:\ Evidence for Excess Emission at 3--8 microns
}
 
\author{Steve B. Howell,\altaffilmark{1} 
Carolyn Brinkworth,\altaffilmark{2} D. W. Hoard,\altaffilmark{2} 
Stefanie Wachter,\altaffilmark{2}
Thomas Harrison,\altaffilmark{3}
Howard Chun,\altaffilmark{4,7} Beth Thomas,\altaffilmark{4,8} 
Linda Stefaniak,\altaffilmark{4,9}
David R. Ciardi,\altaffilmark{5} Paula Szkody,\altaffilmark{6}
Gerard van Belle\altaffilmark{5}
}
\altaffiltext{1}{WIYN Observatory and National Optical Astronomy Observatory,
950 N.\ Cherry Ave., Tucson, AZ 85719 {\it howell@noao.edu}}
\altaffiltext{2}{Spitzer Science Center, California Institute of Technology,
Pasadena, CA 91125}
\altaffiltext{3}{Dept.\ of Astronomy, New Mexico State University,
Box 30001, MSC 4500, Las Cruces, NM 88003}
\altaffiltext{4}{NASA/NOAO Spitzer Space Telescope Observing Program 
for Students and Teachers}
\altaffiltext{5}{Michelson Science Center, California Institute of Technology,
Pasadena, CA 91125}
\altaffiltext{6}{Dept.\ of Astronomy, University of Washington,
Box 351580, Seattle, WA 98195}
\altaffiltext{7}{Cranston High School East, 899 Park Ave., Cranston, RI 
02910}
\altaffiltext{8}{Environmental Education, Great Falls Public Schools,
West Elementary, 1205 1st Ave.\ NW, Great Falls, MT 59404}
\altaffiltext{9}{Allentown High School, 27 High Street, Allentown, NJ 08501}

\begin{abstract}
We present the first observations of magnetic cataclysmic 
variables with the Spitzer Space Telescope. We used the 
Infrared Array Camera to obtain photometry of the polars 
EF Eri, GG Leo, V347 Pav, and RX J0154.0-5947 at 3.6, 4.5, 
5.8, and 8.0 $\mu$m.  In all of our targets, we detect 
excess mid-infrared emission over that expected from the 
component stars alone.  We explore the origin of this IR 
excess by examining bremsstrahlung, cyclotron emission, 
circumbinary dust, and L/T brown dwarf secondary stars.  
Bremsstrahlung and cyclotron emission appear unlikely to be 
significant contributors to the observed fluxes.  At present, 
the most likely candidate for the excess emission 
is dust that is probably located 
in a circumbinary disk with an inner temperature near 800 K.  
However, a simple dust disk plus any reasonable low mass or brown 
dwarf-like secondary star is unable to fully explain the observed 
flux densities in the 3--8 $\mu$m region.
\end{abstract}

\keywords{Stars: individual (EF Eri, GG Leo, V347 Pav, 
RX J0154.0-5947) --- stars: low-mass --- stars: brown dwarfs}

\section{Introduction}

Cataclysmic variables (CVs) are interacting binary stars containing a
white dwarf (WD) primary and a low mass secondary (Warner 1995). 
CVs that contain a highly magnetic WD,
with surface fields ranging from 
10--250 MG, are called polars.  
In a polar, accreted material flows along 
magnetic field lines, does not form a viscous disk, 
and accretes directly onto the WD at/near its magnetic 
pole(s).  Polars are well known to stop and (re)start their mass transfer,
causing changes of $\sim$2--3 (or more) optical magnitudes at apparently 
random intervals.
During low states, the accretion flux disappears and the two component 
stars can often be cleanly observed.
For the past few decades, theoretical models of CV evolution 
(Kolb 1993; Howell et al.\ 2001) 
have predicted that the very shortest orbital period systems 
should contain low mass, brown dwarf-like secondary stars.
More recently, near-IR ($JHK$) observations of the shortest
period CVs (e.g., Harrison et al.\ 2003; Howell et al.\ 2004) 
have provided evidence that they likely do contain secondaries
similar to L or T dwarfs.

We have accomplished an initial survey of four polars 
(EF Eridani, GG Leonis, V347 Pavonis, and RX J0154.0-5947)
using the Infrared Array Camera (IRAC) 
on the Spitzer Space Telescope. 
We specifically chose ultra-short period polars ($P_{\rm orb} \leq 90$ min)
to facilitate studying
the brown dwarf-like secondary stars predicted to be present.
We found that all four of the polars show emission in the 
3--8 $\mu$m region in excess of that produced 
solely by the WD plus an M or L dwarf. 
In this Letter, we briefly discuss our observations and
possible sources of this excess emission.  The best
explanation is cool circumbinary dust that would 
likely dynamically settle into a disk, and/or possibly a T dwarf-like 
secondary star.

\section{Spitzer Infrared Array Camera Observations}

IRAC is a four channel camera that
obtains nearly simultaneous images at 3.6, 4.5, 5.8, and 
8.0 $\mu$m (see Fazio et al.\ 2005).
For our polars, we obtained 30-sec images using the 
Gaussian-5 dither pattern. 
The dithered images were combined and 
flux calibrated with the S12 Spitzer pipeline.
Details of our data reduction and error analysis
procedures will be presented in Brinkworth et al.\ (2006).
Figure 1 shows the spectral energy distributions (SEDs) for our
polars, comprised of our IRAC data (Table 1) and non-simultaneous 2MASS 
$JHK_{\rm s}$ photometry.
We have classified the polars based on the overall
appearance of their $JHK_{\rm s}$ SEDs.  
EF Eri was known to be in a low accretion state during its 2MASS 
observation (Harrison et al.\ 2004), and we infer that V347 Pav was also, 
based on its similar SED.
We suspect that GG Leo and RX J0154 were both in high accretion states 
during their 2MASS observations, 
because of the falling Rayleigh-Jeans (RJ) tail of accretion flux 
that dominates their $JHK_{\rm s}$ SEDs.
We note that the 2MASS bands are sensitive to the mass accretion 
state of the system (high states show a falling RJ tail, low states probably 
begin to reveal the secondary star) while the IRAC bands are red 
enough to be essentially unaffected by these changes.

\begin{figure}
\includegraphics[angle=-90, scale=0.62]{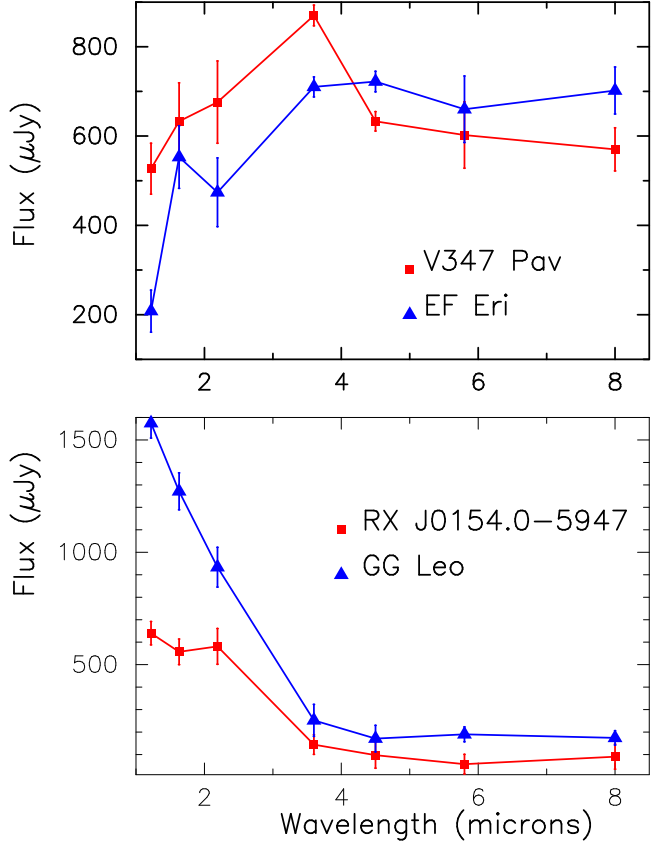}
\caption{Spitzer IRAC photometric observations at 
3.6, 4.5, 5.8, and 8.0 $\mu$m and 2MASS $JHK_{\rm s}$ 
photometry for the polars EF Eri and V347 Pav (top) and 
GG Leo and RX J0154.0-5947 (bottom).}
\end{figure}

%Table 1
\begin{deluxetable}{llllllll}
\tablenum{1}
\tablewidth{6.2in}
\tablecaption{2MASS and Spitzer IRAC Fluxes ($\mu$Jy)}
\tablehead{
   \colhead{Polar}
 & \colhead{1.2 $\mu$m}
 & \colhead{1.6 $\mu$m}
 & \colhead{2.2 $\mu$m}
 & \colhead{3.6 $\mu$m}
 & \colhead{4.5 $\mu$m}
 & \colhead{5.8 $\mu$m}
 & \colhead{8.0 $\mu$m}
}
\startdata
EF Eri                    & \phn208(47) & \phn553(70) & 474(77) & 710(22) &    722(23) &    660(74) &    702(53) \\
V347 Pav                  & \phn527(57) & \phn633(86) & 676(92) & 870(23) &    633(22) &    602(74) &    570(48) \\
GG Leo                    &    1575(66) &    1272(82) & 933(88) & 252(71) &    170(59) &    190(33) &    174(32) \\
RX J0154\tablenotemark{a} & \phn640(52) & \phn557(57) & 581(80) & 144(43) & \phn97(58) & \phn57(43) & \phn91(55) 
\enddata
\tablenotetext{a}{RX J0154 is our faintest target in the IRAC bands; 
however, it was clearly detected in each IRAC channel. Our formal 
1$\sigma$ errors for this star, as listed in the table, are probably 
best taken as lower limits.}
\end{deluxetable}

\section{Source of the Excess 3--8 micron Emission}

Our Spitzer observations show that 
all four polars present a similar, relatively flat SED in the IRAC region.
The 3--8 $\mu$m emission does not fall off to the red as it would if the source
was a hot blackbody or the stellar RJ tail of an M or L star.
We now examine some possibilities for its origin.

\subsection{Cyclotron Emission}

The high magnetic fields in polars can
produce cyclotron emission in the UV--near-IR spectral regions
as accreted material is ionized and the free electrons spiral along field
lines (Warner 1995).
At times, cyclotron cooling can become the dominant cooling
mechanism (Cropper 1990) and its blue spectrum consists of large hump features
$\sim$1000\AA\ wide that can increase the local continuum by up to one
magnitude. 
A thorough review of model cyclotron spectra is given in 
Wickramasinghe \& Ferrario (2000, hereafter WF2000).
Spectral intensity and shape, as well as any harmonic structures (humps), 
are dependent on magnetic field strength, 
electron temperature and density, and viewing angle.
Cyclotron spectra are characterized by a long wavelength RJ tail
and a slowly rising, power-law blue continuum modulated by harmonic
hump features.  The peak of this pseudo-blackbody spectrum 
moves blueward as the magnetic field
strength increases (e.g., Fig.\ 32 in WF2000). 

Fields of 14--30 MG, as found in
our target polars (Warner 1995), produce 
power law continua and observable
humps in the optical--near-IR spectral region,
with optically thick RJ tails at longer wavelengths.
If we (incorrectly) attribute all the $K_{\rm s}$ band flux in EF Eri 
to the RJ tail of cyclotron emission, then
we would expect the IRAC observations to produce
flux densities of $\approx$ 114, 58, and 38 $\mu$Jy  at 4, 6, and 8 $\mu$m, 
respectively.
This is not consistent with our observations.
In order for the cyclotron spectrum alone to produce a relatively 
flat SED in the 3--8 $\mu$m region,
the IRAC observations would have to sample the 
power-law portion (plus humps). 
This would require magnetic field strengths of only a few MG, which 
may be found in polars at large distances from the WD ($B \propto r^{-3}$). 
Fields of 1--2 MG occur at about 2--3 WD radii
in our polars, assuming conservation of magnetic 
flux density.  At the same 
viewing geometry and opacity, the emission volume in this low field 
region will be larger than at the accretion pole by a 
factor $(r/R_{\rm WD})^3$, but the electron temperature and 
material density will be lower.

The ratio of overall spectral intensity from
the low field region ($I_r$) compared to the pole region ($I_R$) 
is proportional to 
$$I_r/I_{R_{\rm WD}} \propto (B_{\rm r}/B_{\rm WD})(r/R_{\rm WD})^3 N_{\rm e} T_{\rm e} e^{-(r/R_{\rm WD})} \approx 0.07,$$ 
(see Eqn.\ 53 in WF2000) where $N_{\rm e} \sim 10^{14}$ cm$^{-3}$ and $T_{\rm e} \sim 5$ keV, 
typical low state values at the magnetic pole 
on the WD surface.
If we assign all the 3.6 $\mu$Jy flux in EF Eri (711 $\mu$Jy) 
to cyclotron emission from the main field, then weak field cyclotron
radiation would produce a flat SED of only $\sim 50$ $\mu$Jy across 
the IRAC bands.  This is much fainter than our Spitzer data, but is in
agreement with recent limits set by non-detection of cyclotron emission 
in Spitzer observations of 
intermediate polars (Johnson et al.\ 2005), which are 
believed to have WD surface fields of a few MG.  
Thus, it seems that any cyclotron emission present in the IRAC 
bands would provide
only a small addition to the observed flux, possibly as a low level RJ tail,
and cannot, in itself, account for
the observed shape or level of the 3--8 $\mu$m SED in our polars.

\subsection{Bremsstrahlung}

Bremsstrahlung, or free-free emission, produces a flat SED 
and can be the dominant continuum source at radio wavelengths 
or in high temperature environments.
The total energy (in ergs) radiated per 
cm$^{3}$ per second for a pure hydrogen environment is given by
$$E_{\rm ff} = 1.4 \times 10^{-27} N_{\rm e}^{2} T^{0.5} \langle g \rangle,$$
where $\langle g \rangle$, the Gaunt factor, is near 1 (Spitzer 1978).
Bremsstrahlung is nearly independent of
frequency until its cutoff frequency, a value set by the electron thermal 
velocity distribution (i.e., $h\nu \approx kT$).
At high temperatures ($>> 10^{6}$ K)
this cutoff frequency is in the optical--IR, while lower
temperatures only produce significant bremsstrahlung 
in the radio.

At typical electron densities ($N_{\rm e} \sim 10^{13}$ cm$^{-3}$) and 
high state temperatures
(100,000--600,000 K) near the WD pole(s) (Warner 1995), and 
assuming an emission region of 0.1 of the WD area, 
$E_{\rm ff} \sim 2600$ erg s$^{-1}$. 
Over the IRAC bands, even such high temperature regions 
would provide a continuum level
of only about $10^{-10}$ $\mu$Jy. This is
$\sim 10^{13}$ less than expected from the RJ tail of a 1500 K
star at the distances of our polars (Rybicki \& Lightman 1979).
Looking instead at low temperature environments within the CV,
a temperature of only $\sim 3600$ K will produce a
cutoff frequency at 4 $\mu$m.  However, the 
number density of free electrons continuously kept at this temperature, 
which is required to produce measurable bremsstrahlung
in the IRAC bands, must exceed $10^{15}$ cm$^{-3}$. 
This is a highly unrealistic value -- it would require the total accumulated 
mass of many thousands of years of
high state mass transfer ($10^{-12} M_{\odot}$ yr$^{-1}$) 
between the two stars.  In order to lower the 
number density of 3600 K material to even a remotely plausible value
of $10^{11}$ cm$^{-3}$, a spherical radiating volume 
with radius near $10^{15}$ cm is needed. This is about 
100,000 times the size of the binary orbit.
We can, therefore, likely eliminate bremsstrahlung as the sole or major
source of the observed IR SEDs in our polars.

\subsection{Circumbinary Dust}

An interesting theoretical concept has been pursued
of late (e.g., Taam \& Spruit 2001) that predicts the existence of 
circumbinary disks of
cool material in CVs. Observational studies to find such disks have provided
mixed results (Belle et al.\ 2004; Dubus et al.\ 2004)
and, to date, there is little direct observational evidence 
for their existence. 

Becklin et al.\ (2005) present observations of a dust disk surrounding 
GD362, a single, cool ($T=9740$ K) WD. 
In Figure 2, we show the SEDs for EF Eri and V347 Pav (from Figure 1) with
the optical--near-IR model of Harrison et al.\ (2003)
plus the Becklin et al.\ model dust disk, scaled to the 
distance of EF Eri (45 pc).
The Harrison et al.\ model for 
EF Eri spans 3500\AA\ to the $K$ band and is based on actual 
observations of the cool ($T=9500$ K)
WD, and near-IR photometry and spectroscopy that led to a 
best-fit secondary approximating the SED of an L8 star.
Note the negligible contribution from the WD in EF Eri seen in the extreme 
lower left corner of Figure 2. This illustrates the steepness of a 
stellar RJ tail.
The Becklin et al.\ dust disk with $T_{inner}=1200$ K is likely to be 
too hot for our polars, as the inner edge of the
circumbinary disk will be farther from the WD. 
Scaling the inner edge of the disk in GD362 to that in our
binaries (assuming the inner edge must be outside the two stars) we find
that T$_{inner}$ would be near 800$\pm$200 K. 
This cooler disk model is also plotted in Figure 2.
The dust disk model does a fairly good job of fitting the polar 
SEDs up to $\sim 5$ $\mu$m but then 
rises too fast toward the red. The cooler 800 K model fits better but is still 
too bright in the longest 
IRAC band. The dust disk scaling performed here, and the
better fit afforded by the cooler disk, are simple ad hoc
models at present. Detailed, more realistic dust disk models are being 
produced and will be presented in Brinkworth et al.\ (2006).

\begin{figure}
\includegraphics[angle=-90, scale=0.345]{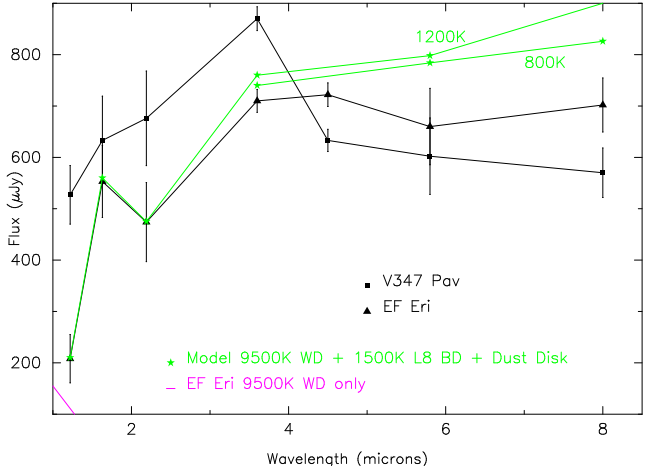}
\caption{The SEDs of EF Eri and V347 Pav from Figure 1
with model SEDs (green lines)
based on optical--near-IR observations of EF Eri plus a 
dust disk based on observations of GD362 (see text for details).  
The WD-only contribution in EF Eri is shown in the extreme lower 
left (purple line).}
\end{figure}

\subsection{L or T Dwarf Secondary Stars}

We compared the SEDs of EF Eri and V347 Pav to those of single
L and T dwarfs from the IRAC observed sample kindly
provided by Patten et al.\ (2006). 
The polar secondary stars have attained their present size, mass, and 
temperature after Gyr of mass loss; hence, they are
brown dwarf-like but are unlikely to be exactly like brown dwarfs.
The SED of an L6 dwarf falls off after 4.5 $\mu$m (dropping by a factor of
2 in brightness from 4.5 to 8 $\mu$m), while the SEDs of late T dwarfs
remain relatively flat throughout the IRAC bands.
We conclude that T dwarf SEDs are more consistent than those of L dwarfs
with our observed spectral shapes in the IRAC bands, but neither type of star 
reproduces the observations in detail.  In particular, T dwarfs are quite
faint, and would require the CVs to be at very small distances in order 
to contribute a large fraction of the observed IRAC flux densities.
For example, if the IRAC fluxes were entirely due to the secondary star, 
EF Eri would have to be at $\approx$12 and $\approx$6 pc if the secondary star
were a T6.0 and T8.0 respectively.

\section{Discussion and Summary}

Figure 3 shows an IRAC two-color diagram for our polars, along with 
several brown dwarfs from the Patten et al.\ sample
(L6 = SDSS 1331-0116; T5 = 2MASS 0559-1404; T8 = 2MASS 0415-0935). 
We have also plotted GD1400, a detached WD+L5/6 
binary with no dust (Farihi et al.\ 2006), and our simple
dust disk model (based on GD362), as shown in Figure 2.
The effect of changing the GD362 model disk temperature 
from $T_{\rm inner}=1200$ K to 800 K is shown in the lower 
right of the figure. 
For comparison, the IRAC colors of T Leonis and VY Aquarii are 
also plotted; 
these are not-yet-published CVs that we have observed with Spitzer. 
They have similar orbital periods and distances as our polars, 
but contain non-magnetic WDs surrounded by hot ($\sim 8000$ K) accretion 
disks.  The SEDs of T Leo and VY Aqr are dominated by the RJ tails of the
secondary star and accretion disk well into the IRAC region.
Our observed polars occupy a region of color-space intermediate between
WD+brown dwarf binaries without dust, single brown dwarfs, and dust 
disk-dominated systems.

\begin{figure}
\includegraphics[angle=-90, scale=0.335]{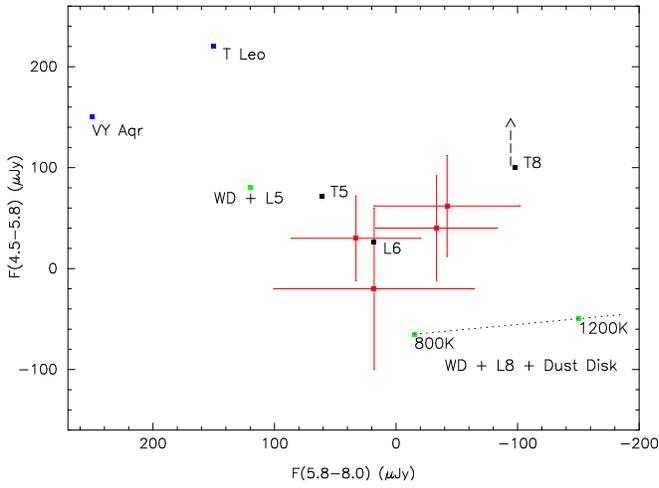}
\caption{IRAC color-color diagram showing our four polars 
(red points with error bars). The WD+L5 green point is GD1400 (Farihi et al.
2006) and the two green points at 800K and 1200K, represent our scaled GD362
(WD+L8) dust disk model as described in \S3.3.
The white points are observed single L6, T5, and T8 dwarfs, and 
the two blue points are observed
accretion disk dominated CVs (T Leo and VY Aqr).
The F(4.5$-$5.8) value for the T8 dwarf 
is off scale at 1384 $\mu$Jy as indicated by the arrow in the figure.
See the text for details.}
\end{figure}

Bremsstrahlung, while often producing a flat spectral
continuum, likely contributes essentially nothing 
to our observed IRAC fluxes.
Cyclotron emission also appears unlikely to be a major contributor
to the SEDs of our polars, since
it should be present only as a low level falling RJ tail.
However, our understanding of the complex nature of cyclotron 
emission is far from
perfect, so we hesitate to say with complete certainty that none of the excess
emission is due to the cyclotron process.
At present, the best candidates for the origin of the observed IR excess 
are cool dust (in or near the
CV) and/or secondary stars that are similar to late T dwarfs.

If the dust is produced by aeons of mass outflow from 
the CV (in winds or nova outbursts) or is possibly remnant material 
from the common envelope stage ejected with some orbital angular momentum
(e.g., Hinkle et al.\ 2006), 
then dynamic considerations lead us to expect that this material would 
settle into a circumbinary
disk. If a dust disk is present and is similar to that seen around GD362,
then we expect the inner disk temperature to be around 800 K. 
The emitting volume of such a disk must be large compared to the
component stars (which are Earth-sized and Jupiter-sized),
as it outshines even the cool brown dwarf-like secondary in the IRAC bands.
If the 3--8 $\mu$m SED originates primarily from the secondary star,
then it would have to be similar to a 
late T dwarf and these CVs would have to be far closer than observations 
of their WDs suggest.

%\lastpagefootnotes

The likely presence of a circumbinary dust disk in polars naturally leads us to 
postulate that such disks might be common in other CVs.
The 3--8 $\mu$m region still
contains a strong spectral contribution from the RJ tail of the 
accretion disk in CVs with weakly- or non-magnetic WDs.
In longer period systems, the
larger, brighter G, K, and early M secondary stars also 
contribute in the IRAC bandpass.
Thus, the contrast from the weak IR emission of a dust disk against that 
produced by 
the stellar and other components in all but the polars may prevent detection
shortward of 8 $\mu$m. 
Some support for this is provided by 
Dubus et al.\ (2004), who detected
SS Cygni and AE Aquarii at 11 and 17 $\mu$m, and by Abada-Simon et al.\ (2005),
who detected AE Aqr at wavelengths out to 170 $\mu$m\footnote{AE Aqr 
is a special case, however, as most of its far-IR emission is
attributed to synchrotron emission.}.

Neither a simple dust disk, as described here, nor a brown dwarf-like
secondary star alone can fully explain the observed IR excess in the polars, 
especially redward of $\approx 6$ $\mu$m. Spectral observations would help 
distinguish between the possibilities discussed above and provide details
regarding the cause of the observed IR excess.

\acknowledgments

This work is based on observations made with the Spitzer Space 
Telescope, which is operated by the Jet Propulsion Laboratory, 
Caltech, under NASA
contracts 1407 and 960785.
We thank the Spitzer Science Center (SSC) Director for his 
generous allocation of observing time for the NASA/NOAO Spitzer 
Space Telescope Observing Program for Students and Teachers. 
NOAO, which is operated by the Association of Universities for Research 
in Astronomy (AURA), Inc., under cooperative agreement with the 
NSF, has provided many in kind contributions for which the 
first author is grateful. 
Michelle Smith and Kimmerlee 
Johnson from Great Falls High School, Great Falls, Montana assisted in
the data processing. 
We also thank the SSC folks for their hospitality 
during our visit. 
The TLRBSE Project is funded by the NSF under ESI 0101982 
through the AURA/NSF Cooperative Agreement AST-9613615. 
This work makes use of data products from the 
Two Micron All Sky Survey, which is a joint project of the 
University of Massachusetts and IPAC/Caltech,
funded by NASA and the NSF. 
CSB acknowledges support from the SSC Enhanced Science Fund, NASA's 
Michelson Science Center, and the Spitzer Director's Discretionary funds.

\end{document}